\documentclass[aps,pra,superscriptaddress,reprint,floatfix,longbibliography,showpacs,amsfonts,amsmath,amssymb]{revtex4-1}
\usepackage[english]{babel}
\usepackage{amsmath}
\usepackage{amssymb}
\usepackage{graphicx}
\usepackage{hyperref}
\usepackage{upgreek}
\usepackage{epstopdf}
\usepackage{color}
\usepackage[note-name=, use-sort-key = false]{notes2bib}

\begin{document}
\title{Infrared spectroscopy study of the nodal-line semimetal candidate ZrSiTe under pressure: Hints for pressure-induced phase transitions}

\author{J. Ebad-Allah}
\affiliation{Experimentalphysik II, Augsburg University, 86159 Augsburg, Germany}
\affiliation{Department of Physics, Tanta University, 31527 Tanta, Egypt}
\author{M. Krottenm\"uller}
\affiliation{Experimentalphysik II, Augsburg University, 86159 Augsburg, Germany}
\author{J. Hu}
\affiliation{Department of Physics, University of Arkansas, Fayetteville, AR 72701, USA}
\author{Y. L. Zhu}
\affiliation{Department of Physics, Pennsylvania State University, University Park, PA 16803, USA}
\affiliation{Department of Physics and Engineering Physics, Tulane University, New Orleans, LA 70118, USA}
\author{Z. Q. Mao}
\affiliation{Department of Physics, Pennsylvania State University, University Park, PA 16803, USA}
\affiliation{Department of Physics and Engineering Physics, Tulane University, New Orleans, LA 70118, USA}
\author{C. A. Kuntscher}
\email{christine.kuntscher@physik.uni-augsburg.de}
\affiliation{Experimentalphysik II, Augsburg University, 86159 Augsburg, Germany}

\begin{abstract}
We studied the effect of external pressure on the optical response of the nodal-line semimetal candidate ZrSiTe by reflectivity measurements. At pressures of a few GPa, the reflectivity, optical conductivity, and loss function are strongly affected in the whole measured frequency range (200-16500~cm$^{-1}$), indicating drastic changes in the electronic band structure. The pressure-induced shift of the electronic bands affects both the intraband and interband transitions. We find anomalies in the pressure dependence of several optical parameters at the pressures $P_{c1}$$\approx$4.1~GPa and $P_{c2}$$\approx$6.5~GPa, suggesting the occurrence of two phase transitions of either structural or electronic type.
\end{abstract}

\maketitle
Topological semimetal materials such as Dirac, Weyl, and nodal line semimetals, are of a great current interest due to their exceptional physical properties, including high bulk carrier mobility and large magnetoresistance  \cite{Burkov.2011,Neupane.2014,Fang.2015,Lv.2016,Sankar.2017,Hu.2017,Armitage.2018}. Over the last few years,  nodal-line semimetals (NLSMs) attracted increasing attention due to their multiple band crossings along a line in momentum space, in contrast to Dirac and Weyl semimetals, where the bands are touching at discrete $k$ points. ZrSiS and its isostructural compounds Zr$X$$Y$ ($X$=Si, Ge, Sn and $Y$= O, S, Se, Te) are NLSM candidates, where electronic band structure calculations showed multiple linear band crossings at various energies \cite{Xu.2015,Schoop.2016,Klemenz.2018}. Zr$X$$Y$ compounds have a PbFCl-type structure (tetragonal P4/$nmm$ space group with nonsymmorphic symmetry), where the Zr atoms are coordinated by four carbon group atoms $X$ and four chalcogen atoms $Y$ \cite{Wang.1995}. The Zr, $X$, and $Y$ atoms are arranged in square nets parallel to the $ab$ plane. Accordingly, the crystal structure of the Zr$X$$Y$ compound is layered, consisting of slabs with five square nets and the stacking sequence [$Y$-Zr-$X$-Zr-$Y$].

Several studies reported that the unique electronic properties of Zr$X$$Y$ compounds are mainly determined by the carbon group $X$ square sublattice and that the structural dimensionality can be tuned by isoelectronic substitution of either chalcogen element $Y$ or
the carbon group element $X$ \cite{Wang.1995,Schoop.2016,Ali.2016,Singha.2017}. For example,
tuning the chalcogen atom from S to Te increases the ionic radius
and the ratio of lattice parameters $c/a$ , concomitant with the decrease of the interlayer bonding \cite{Klein.1964,Xu.2015,Topp.2016}.
Accordingly, ZrSiTe is expected to be more two-dimensional (2D) as compared to ZrSiS regarding its electronic properties \cite{Wang.1995}. Indeed, de Haas-van Alphen quantum measurements revealed a 2D character of the Fermi surface (FS), in addition to a 3D component, in contrast to ZrSiS (as well as ZrSiSe) with a 3D-like FS \cite{Hu.2016}.

Within the Zr$X$$Y$ compound family, ZrSiS is the most studied one. ZrSiS has an almost ideal nodal-line electronic band structure with a
diamond-rod-shaped FS, although spin-orbit coupling opens up a small energy gap ($\sim$0.02~eV) at the Dirac nodes     \cite{Neupane.2016,Schoop.2016,Topp.2016}. Additional Dirac-like band crossings, which are protected by non-symmorphic symmetry against gapping (called non-symmorphic band crossings in the following), are located several hundred meV above and below the Fermi energy at the $X$ and $R$ point of the Brillouin zone. These non-symmorphic band crossings were suggested to host novel 2D Dirac fermions with unusual electronic properties \cite{Schoop.2016,Young.2015}.
Recent density functional theory calculations \cite{Topp.2016} showed that such band crossings are also present in other Zr$X$$Y$ compounds, and that their energy position depends on the $c/a$ ratio, i.e., the dimensionality of the system: For compounds with $c/a$ ratios between 2.2 and 2.3, as in ZrSiS, the non-symmorphic band crossings are located $\sim$0.6~eV above and below the Fermi energy ($E_F$), with an energy separation $\Delta$$E_{bc}$$\sim$1.2~eV.
For ZrSiTe with a much larger $c/a$ ratio ($\approx$ 2.57) and corresponding enhanced two dimensionality, these band crossings are shifted towards $E_F$ with $\Delta$$E_{bc}$$\sim$0.4~eV, thus suggesting ZrSiTe to be the first real material that exhibits non-symmorphic band crossings close to $E_F$ \cite{Topp.2016}.

The near-$E_F$ band structure of ZrSiTe thus contains the linear crossing bands of the nodal line, although in comparison to ZrSiS slightly shifted in energy and more gapped due to the larger spin-orbit coupling.
Additionally, there are non-symmorphic band crossings close to $E_F$ in the vicinity of the $X$-$R$ line, which distort the nodal-line structure along the $X$-$R$ line, and the Fermi level is pushed away from the nodal line in the rest of the Brillouin zone \cite{Ebad-Allah.2019}.
It was furthermore shown that ZrSiTe has a diamond-shaped FS, similar to ZrSiS, but with extra small pockets, most probably of hole-like nature \cite{Hosen.2017}.

\begin{figure*}[ptb]
\includegraphics[width=1\textwidth]{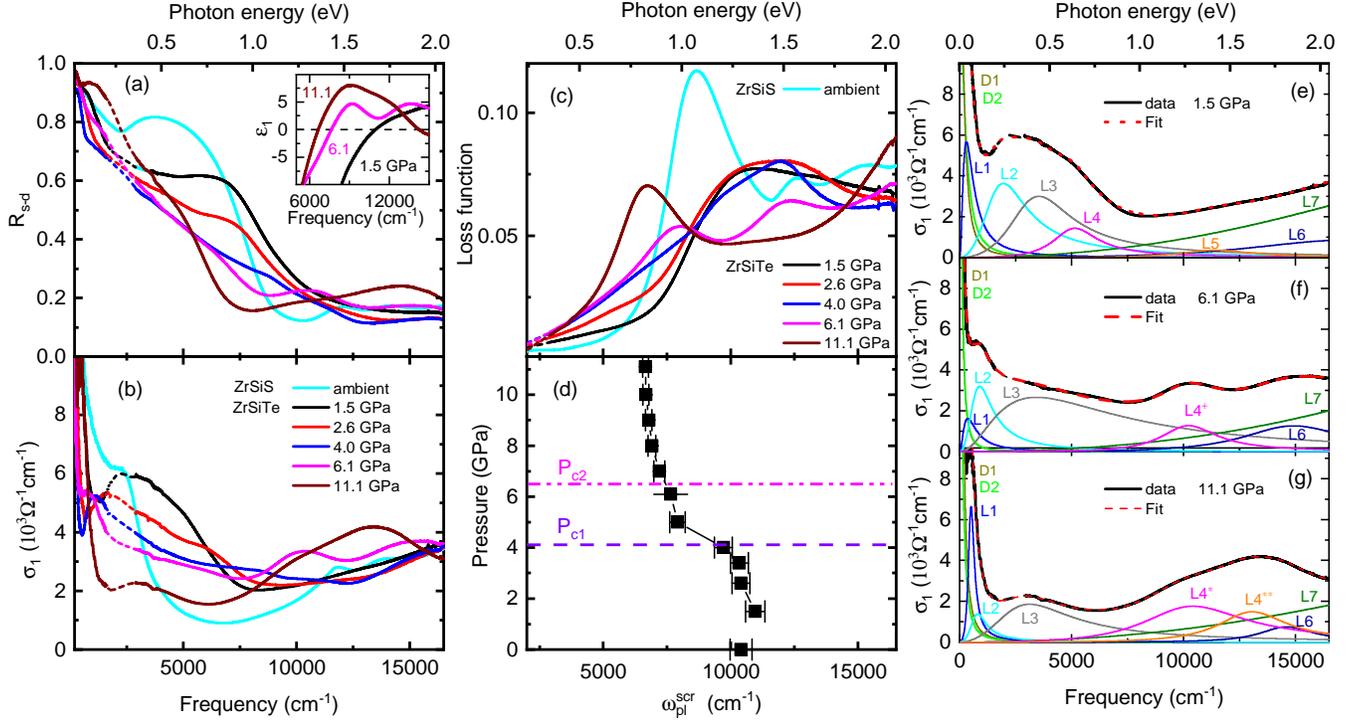}
\caption{Room-temperature optical response functions of ZrSiTe for selected pressures, compared to ambient-pressure ZrSiS: (a) reflectivity R$_{s-d}$ at the sample-diamond interface, (b) real part of the optical conductivity $\sigma_1$, (c) loss function -Im(1/${\hat{\epsilon}}$). The part of the optical response functions, which is affected by the multi-phonon absorption in the diamond, is indicated by a dashed line.
The inset of (a): Real part of the dielectric function $\epsilon_1(\omega)$ of ZrSiTe at 1.5, 6.1, and 11.1~GPa.
(d) Screened plasma frequency $\omega_{pl}^{scr}$ of ZrSiTe as a function of pressure, determined from the zero-crossing energy of $\epsilon_1(\omega)$. The pressure $P_{c1}$ ($P_{c2}$) is marked by a dashed (dash-dotted) line.
(e)-(g): Optical conductivity spectra of ZrSiTe at 1.5, 6.1, and 11.1~GPa, resp., together with the total fit and the various Drude (D) and Lorentz (L) contributions.}
\label{figure1}
\end{figure*}

The application of external pressure is a superior way for tuning the dimensionality of a system, as compared to isoelectronic substitution (chemical pressure). In ZrSiS hydrostatic pressure has the strongest effect on the $c$ lattice parameter, which is perpendicular to the square lattice plane, resulting in a pressure-induced decrease of the
$c$/$a$ ratio \cite{Singha.2018}. In analogy, applying hydrostatic pressure on ZrSiTe, which is more 2D as compared to ZrSiS, a dimensionality change from 2D towards more 3D is expected, concomitant with the decrease of the $c$/$a$ ratio.
According to Topp et al.\ \cite{Topp.2016}, the energy separation $\Delta$$E_{bc}$ of the non-symmorphic band crossings at the $X$ point of the Brillouin zone increases with decreasing $c$/$a$ ratio. Thus, ZrSiTe is expected to have an increased $\Delta$$E_{bc}$ under pressure.

Indeed, in our infrared spectroscopy study of ZrSiTe under external pressure, we find indications of such changes in the electronic band structure at low pressures. The optical response functions are strongly affected by the pressure application.
In particular, we observe anomalies in several optical parameters, signalling the occurrence of two pressure-induced phase transitions.

The optical response functions of ZrSiTe at the lowest studied pressure (1.5~GPa), depicted in Fig.\ \ref{figure1} \cite{Suppl}, are in good agreement with the ambient-pressure results reported recently \cite{Ebad-Allah.2019}. We show here the reflectivity R$_{s-d}$ measured at the sample-diamond interface, the real part of the optical conductivity $\sigma_1$, and the loss function, defined as -Im(1/${\hat{\epsilon}}$) where $\hat{\epsilon}$ is the complex dielectric function (see Supplemental Material for details on experiment and analysis of data).
The reflectivity R$_{s-d}$ is high at low frequencies, indicating the metallic state, but drops with increasing frequency and saturates above 2500~cm$^{-1}$, forming a plateau in the range 2500-7000~cm$^{-1}$ [see Fig.\ \ref{figure1}(a)]. The further rapid decrease of reflectivity above $\approx$7000~cm$^{-1}$ marks the onset of the plasma edge, which is followed by a feature-less, low reflectivity above $\approx$12,000~cm$^{-1}$. The plasma edge in ZrSiTe at 1.5~GPa is rather broad and not well defined because of the plateau. This is also revealed by the loss function, which does not contain a clear plasmon mode [Fig.\ \ref{figure1}(c)]. In contrast, for ZrSiS at ambient pressure the reflectivity spectrum shows a rather sharp plasma edge [Fig.\ \ref{figure1}(a)], and the corresponding loss function [Fig.\ \ref{figure1}(c)] contains a well-defined peak at the screened plasma frequency $\omega_{pl}^{scr}$$\approx$8650~cm$^{-1}$, which corresponds to the intraband plasmon.
Alternatively, $\omega_{pl}^{scr}$ can be determined from the zero-crossing of the real part of the dielectric function, $\epsilon_1(\omega)$ \cite{Wooten.1972}. According to the zero-crossing of $\epsilon_1(\omega)$ [inset of Fig.\ \ref{figure1}(a)], we obtain $\omega_{pl}^{scr}$=11000~cm$^{-1}$ for ZrSiTe at 1.5~GPa.

The optical conductivity $\sigma_1$ for ZrSiTe at 1.5~GPa contains Drude contributions at low frequencies related to the itinerant charge carriers and a broad absorption band centered at around 3200~cm$^{-1}$ due to excitations between electronic bands close to $E_F$
[see Fig.\ \ref{figure1}(b)]. Above 8000~cm$^{-1}$ the optical conductivity monotonically increases with increasing frequency which signals the onset of higher-energy interband transitions.
The contributions to the optical conductivity at 1.5~GPa, as obtained by Drude-Lorentz fitting, are depicted in Fig.\ \ref{figure1}(e) (see also Supplemental Material):
The low-energy optical conductivity is described by two Drude terms D1 and D2 and one sharp Lorentzian term L1.
The two Drude contributions in ZrSiTe are consistent with the presence of electron- and hole-type itinerant charge carriers, with the prevalence of the latter according to magnetotransport measurements \cite{Hosen.2017}.
From the spectral weight of the Drude contributions we obtain the plasma frequency
$\omega _{pl}$=20.000$\pm$1400~cm$^{-1}$
according to the sum rule $\omega _{pl}^{2}/8$=$\int_{0}^{\infty}(\sigma _{1,D1}(\omega) + \sigma _{1,D2}(\omega)) d\omega$ \cite{Wooten.1972}
\bibnote{The lowest-energy Lorentz term L1 strongly overlaps with the Drude terms D1 and D2, and is therefore hard to distinguish. Taking into account the spectral weight of L1 in the sum rule according to
$\omega _{pl}^{2}/8$=$\int_{0}^{\infty}(\sigma _{1,D1}(\omega) + \sigma _{1,D2}(\omega) + \sigma _{1,L1}(\omega)) d\omega$, gives the plasma frequency $\omega _{pl}$=24730~cm$^{-1}$.
It is important to note that the anomalies in $\omega _{pl}$ as a function of pressure are observed independent of whether the spectral weight of the Lorentz term L1 is taken into account in the sum rule or not.}.
The plasma frequency $\omega_{pl}$ is related to the screened plasma frequency
$\omega_{pl}^{scr}$ according to $\omega_{pl}^{scr}$=$\omega_{pl}$/$\sqrt{\epsilon_{\infty}}$, where $\epsilon_{\infty}$ is the high-frequency value of $\epsilon_1(\omega)$. Hence, one obtains $\epsilon_{\infty}$$\approx$3.3, in good agreement with $\epsilon_1(\omega)$ depicted in the inset of Fig.\ \ref{figure1}(a).
The broad absorption band centered at $\sim$3200~cm$^{-1}$ is fitted with three Lorentz terms labeled L2 -- L4. The optical conductivity above 9000~cm$^{-1}$ is described by two Lorentzians (L5 and L6) together with a broad, high-energy background (L7) due to higher-energy excitations, which is present for all pressures and will not be discussed in the following.


The profile of the optical conductivity of ZrSiTe at ambient/low pressure is distinctly different from that of other Zr$XY$ materials, whose optical conductivity has a characteristic U shape \cite{Ebad-Allah.2019,Schilling.2017}.  As an example, we include in Fig.\ \ref{figure1}(b) the $\sigma_1$ spectrum of ZrSiS at ambient pressure.
The low-energy range ($<$6000~cm$^{-1}$) of the U-shaped optical conductivity stems from transitions between linearly crossing bands
along a surface in the Brillouin zone, forming an effective nodal plane \cite{Ebad-Allah.2019}.
In contrast, the optical conductivity of ZrSiTe does not show a U shape, but the low-energy region is dominated by a broad absorption band at $\sim$3200~cm$^{-1}$ ($\sim$0.4~eV), which we attribute to transitions between electronic bands close to $E_F$.
According to theoretical calculations, the near-$E_F$ electronic band structure of ZrSiTe contains the linearly dispersing bands of the nodal line, like in ZrSiS, but with a larger energy gap due to the larger spin-orbit coupling coupling (energy gap size $\sim$60~meV in ZrSiTe as compared to $\sim$20~meV in ZrSiS) \cite{Neupane.2016,Schoop.2016,Topp.2016,Hosen.2017}.
Additionally, in ZrSiTe non-symmorphic band crossings appear close to $E_F$, with an energy
separation $\Delta$$E_{bc}$$\approx$0.4~eV at the $X$ point of the Brillouin zone \cite{Topp.2016}.
These band crossings also contribute to the low-energy interband transitions in ZrSiTe \cite{Ebad-Allah.2019}, though it is difficult to relate them to one specific contribution (L2--L4) of the broad absorption band.

\begin{figure}[t]
\includegraphics[width=0.4\textwidth]{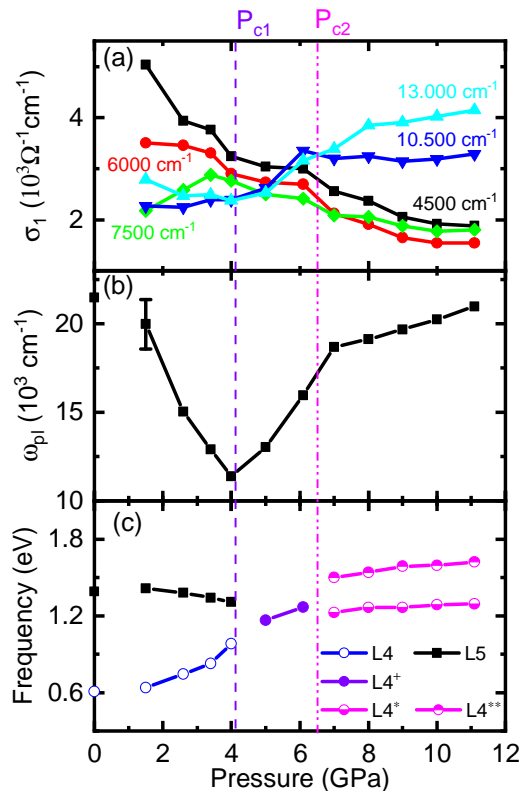}
\caption{Pressure-dependence of (a) $\sigma_{1}$ at various frequencies and (b) plasma frequency $\omega _{pl}$ obtained from the spectral
weight of the Drude terms D1 and D2. (c) Pressure-dependent frequency positions of the contributions L4 and L5 below $P_{c1}$ and their evolution at higher pressures (L4$^{\dagger}$, L4$^{\ast}$, L4$^{\ast\ast}$; see text for more details). The dashed and dashed-dotted vertical lines mark the pressures $P_{c1}$ and $P_{c2}$, respectively.}
\label{figure2}
\end{figure}

With increasing pressure, the optical response of ZrSiTe changes drastically over the whole studied energy range (see Fig.\ \ref{figure1} and Supplemental Material).
For pressures up to $P_{c1}$$\approx$4.1~GPa the overall reflectivity decreases and the plateau in the frequency range 2500 - 7000~cm$^{-1}$ disappears. The two Drude terms in the low-energy optical conductivity lose spectral weight and sharpen. Most interestingly, the absorption band at $\sim$3200~cm$^{-1}$ undergoes major pressure-induced changes: Its contributions L2 and L3 shift to lower energies, whereas its high-energy shoulder (contribution L4) shifts to higher energies [see Fig.\ \ref{figure2}(c)]. The L4 term might be related to transitions involving the non-symmorphic band crossings, since these are
expected to shift away from $E_F$ with increasing pressure (decreasing $c$/$a$ ratio), with a concomitant increase in $\Delta$$E_{bc}$ \cite{Topp.2016}.
Besides, the contributions L3 and L4 broaden and L2 sharpens under pressure, whereas the term L5 slightly shifts to lower energies, loses spectral weight, and disappears at $P_{c1}$ (see Fig.\ \ref{figure2}(c) and Supplemental Material for more details).

For pressures between $P_{c1}$ and 6.1~GPa the low-energy ($\leq$5000~cm$^{-1}$) reflectivity increases with increasing pressure [see
Fig.\ \ref{figure1}(a)], sinalling an increasing metallic character. The low-energy profile of $\sigma_1$ is dominated by a rather sharp L2 peak and a broad L3 contribution, besides
the two Drude terms D1 and D2, whose spectral weight shows a pressure-induced increase as revealed by the increasing $\omega_{pl}$ [Fig.\ \ref{figure2}(b)]. At around 10.000~cm$^{-1}$ a contribution L4$^{\dagger}$ appears, and L6 shifts into the measured frequency range. For illustration, we show in Fig.\ \ref{figure1}(f) the optical conductivity spectrum at 6.1~GPa together with the contributions obtained from the Drude-Lorentz fitting.

At $P_{c2}$$\approx$6.5~GPa a splitting of the L4$^{\dagger}$ peak into two terms L4$^{\ast}$ and L4$^{\ast\ast}$ occurs. Above $P_{c2}$ the increase of the low-frequency reflectivity continues. A more well-defined plasma edge forms [see Fig.\ \ref{figure1}(a)], which is manifested in the loss function by a plasmon mode, located at $\sim$6700~cm$^{-1}$ for 11.1~GPa [see Fig.\ \ref{figure1}(c)].
The profile of the optical conductivity at high pressures is markedly different from the low-pressure regimes [see Fig.\ \ref{figure1}(g)]: The low-energy range is dominated by two Drude terms and the two sharp low-energy Lorentzians L1 and L2, followed by a broad and less pronounced excitation L3. The higher-energy ($\geq$7000~cm$^{-1}$) optical conductivity contains the two Lorentz terms L4$^{\ast}$ and L4$^{\ast\ast}$, which slightly shift to higher energies and gain spectral weight with increasing pressure, and the term L6.

The abrupt pressure-induced changes in the optical response of ZrSiTe at $P_{c1}$$\approx$4.1~GPa and $P_{c2}$$\approx$6.5~GPa are manifested by anomalies in the pressure dependence of several optical parameters, namely in
(i) the screened plasma frequency $\omega_{pl}^{scr}$ [Fig.\ \ref{figure1}(d)], (ii) $\sigma_{1}$ at various frequencies [Fig.\ \ref{figure2}(a)], and (iii) the plasma frequency $\omega _{pl}$ as obtained from the spectral weight of the Drude terms D1 and D2 [Fig.\ \ref{figure2}(b)]. These anomalies suggest the occurrence of phase transitions - either structural or electronic - at the critical pressures $P_{c1}$ and $P_{c2}$.
Apparently, both intraband and interband transitions abruptly change at $P_{c1}$ and $P_{c2}$, suggesting that the phase transitions strongly affect the electronic band structure for energies -1~eV$\leq$$E_F$$\leq$1~eV and thereby the FS.

Overall, the observed drastic changes in the optical properties show that the electronic band structure of ZrSiTe is highly sensitive to external pressure, whereby ZrSiTe remains metallic at all studied pressures. Our results indicate that pressure pushes some electronic bands towards the Fermi level, wheras others are shifted away from $E_F$, with effects on both the intraband as well as the interband transitions. The former is manifested by strong changes in $\omega_{pl}$, wheras the latter by (i) additional contributions to $\epsilon_1(\omega)$ in the higher-energy range (above $\sim$7000~cm$^{-1}$), shifting the zero-crossing of $\epsilon_1(\omega)$ to lower energies concomitant with a decrease in $\omega_{pl}^{scr}$ over the whole studied pressure range, in contrast to the behavior of $\omega_{pl}$, and (ii) redistribution of spectral weight over the whole energy range (see also Fig.\ S3 in the Supplemental Material).
It is important to note, that for all pressures, the profile of the optical conductivity of ZrSiTe does not resemble that of ZrSiS [see Fig.\  \ref{figure1}(b)]. In particular, the U shape of the optical conductivity, which is characteristic for several Zr$X$$Y$ compounds \cite{Ebad-Allah.2019}, is not observed for ZrSiTe at any pressure.
Accordingly, pressurized ZrSiTe seems to be quite different from ambient-pressure ZrSiS regarding its electronic band structure, although a significant decrease of the $c/a$ ratio of ZrSiTe under pressure is to be expected.

We furthermore note that for ZrSiS, with a much reduced $c/a$ ratio at ambient conditions, the occurrence of a pressure-induced topological quantum phase transition for rather low pressures between 0.16 - 0.5~GPa was recently proposed based on Shubnikov-de Haas measurements \cite{VanGennep.2019}.
In another topological material with a layered crystal structure, ZrTe$_5$, the variation of the interlayer interaction was suggested as a possible reason for the observed topological phase transition induced by temperature \cite{Zhang.2017,Xu.2018}.
For a detailed interpretation of the pressure-dependent optical response, information on the crystal structure and electronic band structure of ZrSiTe under pressure is needed.


In conclusion, the strong effects of external pressure on the optical response functions (reflectivity, optical conductivity, loss function) in the whole studied frequency range show that the electronic band structure of ZrSiTe is highly sensitive to pressures. Our observations of the pressure dependence of plasma frequency and optical conductivity indicate that under pressure some electronic bands are pushed towards $E_F$, whereas others shift away from $E_F$, affecting both intraband and interband transitions. Several optical parameters show anomalies in their pressure dependence at  $P_{c1}$$\approx$4.1~GPa and $P_{c2}$$\approx$6.5~GPa, suggesting the occurrence of two phase transitions of either structural or electronic type.

\begin{acknowledgments}
C.A.K. acknowledges financial support from the Deutsche Forschungsgemeinschaft (DFG), Germany, through grant no.\ KU 1432/13-1. The sample synthesis and characterization efforts were supported by the US Department of Energy under grant DE-SC0019068.
\end{acknowledgments}


\begin{thebibliography}{28}
\bibitem [{1()}]{1}%
  \BibitemOpen
  \href@noop {} {}\bibinfo {note} {The lowest-energy Lorentz term L1 strongly
  overlaps with the Drude terms D1 and D2, and is therefore hard to
  distinguish. Taking into account the spectral weight of L1 in the sum rule
  according to $\omega _{pl}^{2}/8$=$\int _{0}^{\infty }(\sigma _{1,D1}(\omega
  ) + \sigma _{1,D2}(\omega ) + \sigma _{1,L1}(\omega )) d\omega $, gives the
  plasma frequency $\omega _{pl}$=24730~cm$^{-1}$. It is important to note that
  the anomalies in $\omega _{pl}$ as a function of pressure are observed
  independent of whether or not the spectral weight of the Lorentz term L1 is taken
  into account in the sum rule.}\BibitemShut {Stop}%
\end{thebibliography}
\end{document}